\documentclass{article}
\usepackage[a4paper, total={5in,8in}]{geometry}
\usepackage{url,hyperref,lineno,microtype,subcaption}
\usepackage[onehalfspacing]{setspace}
\usepackage{wrapfig}
\usepackage{hyphenat}
\usepackage{amsmath}
\usepackage{natbib}
\usepackage{authblk}
\usepackage{graphicx}
\newcommand{\vect}[1]{\boldsymbol{#1}}
\title{Neuronal Sequence Models for Bayesian Online Inference}
\author{Sascha Fr\"olich\thanks{sascha.froelich@tu-dresden.de}}
\author{Dimitrije Markovi\'c}
\author{Stefan J. Kiebel}
\affil{Psychology Department,\\ Chair of Neuroimaging,\\\ Technische Universit\"at Dresden}

\date{}

\begin{document}
\onecolumn

\maketitle
\begin{abstract}
\noindent Sequential neuronal activity underlies a wide range of processes in the brain. Neuroscientific evidence for neuronal sequences has been reported in domains as diverse as perception, motor control, speech, spatial navigation and memory. Consequently,  different dynamical principles have been proposed as possible sequence-generating mechanisms. Combining experimental findings with computational concepts like the Bayesian brain hypothesis and predictive coding leads to the interesting possibility that predictive and inferential processes in the brain are grounded on generative processes which maintain a sequential structure. While probabilistic inference about ongoing sequences is a useful computational model for both the analysis of neuroscientific data and a wide range of problems in artificial recognition and motor control, research on the subject is relatively scarce and distributed over different fields in the neurosciences. Here we review key findings about neuronal sequences and relate these to the concept of online inference on sequences as a model of sensory-motor processing and recognition. We propose that describing sequential neuronal activity as an expression of probabilistic inference over sequences may lead to novel perspectives on brain function. Importantly, it is promising to translate the key idea of probabilistic inference on sequences to machine learning, in order to address challenges in the real-time recognition of speech and human motion. \\

\tiny
{{\bf \noindent Keywords:} neuronal sequences, recurrent neural networks, Bayesian inference, predictive coding, hierarchies of time scales, review} 
\end{abstract}

\section{Introduction}
\subsection{Sequences in the Brain and Around Us}
The world around us and our everyday activities in it are sequentially structured. For example, the morning routine of making a coffee can be divided into sequential subroutines of grinding the coffee beans, putting them into the machine and boiling water. As our own and other people's movements are sequentially structured, the continuous stream of information we receive is of a sequential structure as well. This is for example evident for auditory input, where speech is based on sequential alterations of frequencies and amplitudes, which in turn are created by the sequential motion of the vocal tract. In the visual domain, the same principle applies where someone else's movements are inferred by considering their sequential, temporal structure, in addition to spatial information like posture \citep{giese2003cognitive}.

\begin{figure}[htbp]
\begin{center}
\includegraphics[width=13cm]{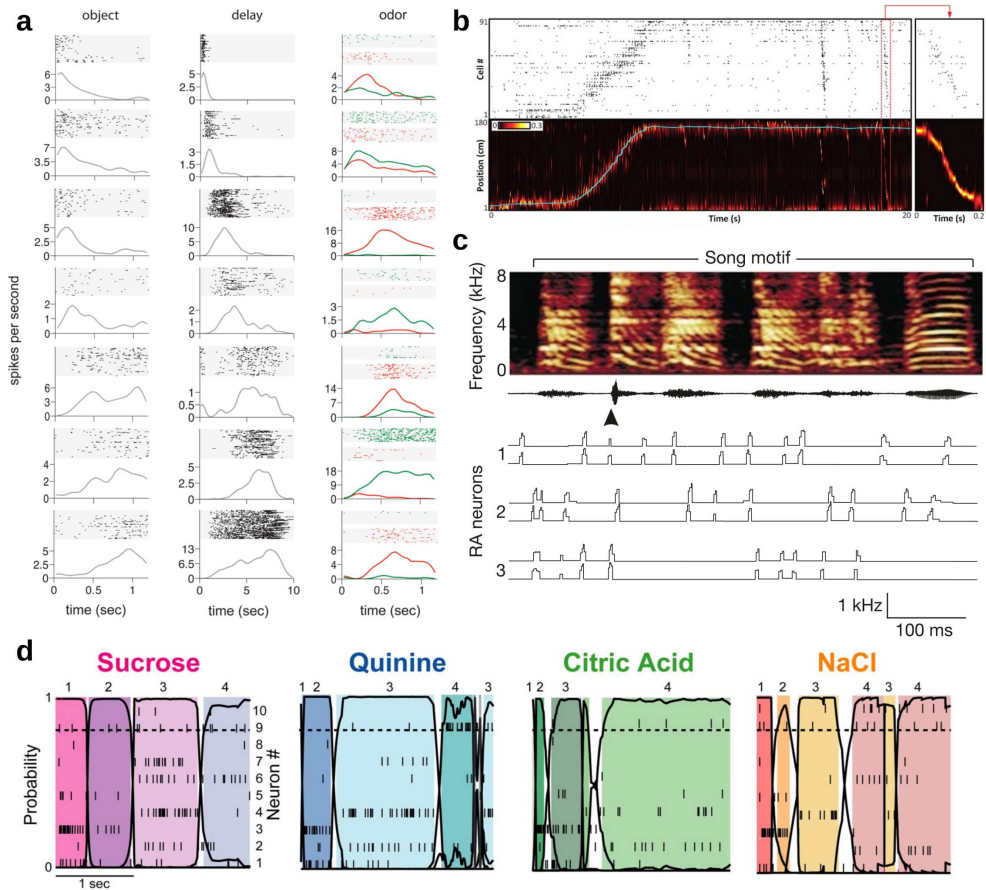}
\end{center}
\caption{Four illustrative examples of sequential neuronal activity in different paradigms and experimental contexts. \textbf{a} Neuronal sequences in the rat hippocampus during different phases of a behavioural experiment. Each row shows the raster plots and peristimulus spike histograms of a different neuron. During the experiment, cells fired sequentially and responded differentially to GO trials (green) and NoGO trials (red). Reprinted from \citep{macdonald2011hippocampal} with permission from Elsevier. \textbf{b} Sequential activation of rat hippocampal cells are not only found during the relevant action but also during rest phases after the behavioural tasks. The top plot shows the spiking histogram of 91 hippocampal cells during a rat's trip along a physical track. The bottom panel shows the rat's actual position on the track (blue line) against the position inferred from the spiking pattern of its hippocampal cells. After the traversal of the track, hippocampal cells ``replayed'' their activation sequence in reverse during a short ripple (red box, enlarged in the box on the right). Figure adapted with permission from \citep{pfeiffer2020content} (Copyright 1999-2019 John Wiley \& Sons, Inc) under the \href{https://creativecommons.org/licenses/by/4.0/legalcode}{Creative Commons License}. (\href{https://dx.doi.org/10.1002/hipo.22824}{DOI}). \textbf{c} Zebra finches are songbirds whose songs consist of highly consistent so-called song motifs. Here, the activation of three different neurons in the RA nucleus of the zebra finch brain during two renditions of the same song motifs is shown. Adapted by permission from \href{https://www.nature.com/}{Springer Nature} \citep{hahnloser2002ultra}. \textbf{d} Firing patterns of neurons in the gustatory cortex of rats \textit{in vivo} when presented with four different odours. While the firing of any single neuron was not predictive of the odour, the sequential switching of states of a Hidden Markov Model (HMM, see section \ref{sec:sequences_in_ML}) was determined by the presented aroma. For each of the four odours, the different colour hues represent different HMM states that were inferred based on the data. Adapted with permission from \citep{jones2007natural} (Copyright (2007) National Academy of Sciences, U.S.A.). }\label{macdonald}
\end{figure}

Given that both our behaviour and the environment are sequentially structured, it is not surprising that sequences of neuronal activity have been reported in a wide range of experimental contexts and seem to play an important role for various brain functions: To date, the brain structure whose functional dependence on sequences is most evidently established is the hippocampus, see \citep{bhalla2019dendrites} for a recent review. For example, hippocampal sequences in a memory task with rats were found to be predictive of future behaviour \citep{pastalkova2008internally}, and sequences in the hippocampus appear to encode and retain context-specific information \citep{macdonald2011hippocampal} (Fig. \ref{macdonald} \textbf{a}). Sequences of hippocampal place cells encoding trajectories in physical space were observed during rest and sleep periods after the relevant spatial experience \citep{skaggs1996replay} (see Fig. \ref{macdonald} \textbf{b}), and similar observations were made during rest before the task \citep{dragoi2011preplay}.

In other areas of the neurosciences, a mesmerising collection of findings has been published in recent years, all of which point at sequential neuronal activity underlying a wide range of distinct perceptual and cognitive processes: For example, sequential activation patterns in working memory tasks have been observed in a number of different brain regions, e.g. the parietal \citep{harvey2012choice} and prefrontal cortex \citep{baeg2003dynamics}. Neuronal sequence detectors were identified in the somatosensory cortex \citep{laboy2019elementary}, and sequential activation of neurons in the gustatory cortex of rats were observed to be involved in the representation of taste \citep{jones2007natural} (Fig. \ref{macdonald} \textbf{d}). There is also evidence for spontaneous neuronal sequences in the visual cortex \citep{kenet2003spontaneously}. In songbirds, temporally highly precise sequences of neuronal firing have been observed during song production and sleep in the so-called HVC and RA regions of the bird brain \citep{hahnloser2002ultra} (Fig. \ref{macdonald} \textbf{c}). In a delayed localisation task, neurons in the frontal cortex of monkeys were found to go through a series of discrete states in a sequence which was predictive of the monkeys' responses \citep{seidemann1996simultaneously}. Recently, spike sequences in the amygdala were found to encode valence-related information during affective learning \citep{reitich2019affective}. Even at the cellular level, there is evidence of sequence processing capacities of single neurons \citep{branco2010dendritic}.

These are a few examples but given the apparent ubiquity of sequential neuronal activity, the question about the potential functional role of neuronal sequences arises. Is there some underlying principle which could unify the functional interpretation of sequential activity in different brain areas? One intuitive explanation is that sequential activity may form a basis for the temporal representation of past, present and future, and is therefore useful to encode the spatio-temporal structure of sensory stimuli \citep{friston2016functional,buonomano2009state}. In particular, theoretical frameworks like predictive coding and the Bayesian brain hypothesis posit that one main function of the brain is to predict future states \citep{friston2009predictive,knill2004bayesian,rao1999predictive}. Consequently, one intriguing idea is that the brain's internal representation and especially prediction relies on sequential representations implemented by neuronal activity \citep{fitzgerald2017sequential,kiebel2009recognizing,hawkins2009sequence}. To infer the hidden causes of sensations, generate predictions and adaptive motor responses, the brain should use a sequential structure similar to the causal structure of the sensory input.

Assuming that sequences are a good basis for entertaining predictions about what happens next in our environment, the next question is what sequential computational mechanism the brain uses to makes sense of the world and generate predictions and actions.

To address this question, we will give a short introduction to the core concepts underlying the brain's function as a predictive organ. This is followed by a discussion on how sequences are typically modelled in the neurosciences and how sequences and hierarchies of sequences are used in both the neurosciences and machine learning. We conclude with a brief outlook on how the principles described here may be useful for converging research in the neurosciences and machine learning.

\subsection{The predictive brain}
Until the early 2000s, the main body of neuroscience research focused on specific functions and faculties of the brain, e.g., memory, attention and visual perception. The  prevailing aim was to find out how the brain implemented these functions. This was followed by a 'predictive revolution' in cognitive neuroscience, where key publications like \citep{friston2005theory, rao1999predictive, knill2004bayesian} would inspire cognitive neuroscientists to integrate the obvious need of the brain to make predictions with specific research questions, e.g. \citep{bubic2010prediction, buchel2014placebo}.      

Importantly, accurate predictions require continuous updating of the causal model of sensory stimuli. Combining the Bayesian brain hypothesis with predictive coding provides a theoretical basis for computational mechanisms that drive a lifelong learning of the hierarchical causal model of the world \citep{friston2014computational}. Under the Bayesian view the neuronal activity does not only encode predictions (and prediction errors) about the hidden causes of sensory stimuli, but additionally encodes a measure of uncertainty about that information \citep{knill2004bayesian}. In order to infer the hidden states of its environment, i.e. to perform recognition, an agent evaluates previously acquired information and incoming sensory data on the basis of an internal \textit{generative model} of itself and its surroundings. On the basis of this model, the agent's experience is consolidated into a \textit{prior distribution}, or \textit{prior belief}, and incoming sensory information informs the so-called \textit{likelihood}. \textit{Bayesian inference} describes how prior and likelihood are combined in an optimal manner to create the \textit{posterior distribution}, which is the agent’s belief about the hidden states of the world \citep{knill2004bayesian}.

Implementations of the Bayesian brain hypothesis have been used to explain, among other things, movement planning \citep{friston2015active} and the computational origins of brain dysfunctions such as psychosis \citep{adams2013computational, corlett2009drugs}, and schizophrenia \citep{fletcher2009perceiving}. See also \citep{clark2013whatever} for a review. Importantly, the Bayesian brain hypothesis \citep{doya2007bayesian} receives experimental support by a wide range of behavioural findings which show that humans tend to act Bayes-optimally, as in, for example, multi-sensory perception \citep{rohe2015cortical, kayser2015multisensory}, spatial and temporal perception \citep{o2013brain,shi2013bayesian}, adaptive learning \citep{soltani2019adaptive, moens2019learning}, perceptual and goal-directed decision making \citep{friston2014anatomy,bitzer2014perceptual,meyniel2015confidence,poudel2017neural}. 

The quality of the inference, that is, the quality of the belief about the hidden states of the world, is dependent on the quality of the agent’'s generative model, and the appropriateness of a tractable (approximate) inference scheme. In this review paper, we suggest that one hallmark of these models in our typical environment is their sequential structure, and that such a sequential structure can be used to robustly perform tractable inference on sequentially structured data.

\subsection{What are sequences?}
\label{sec:what_are_sequences}
Historically, neuroscience research was focused less on the temporal domain but on more obvious, accessible research questions, like the brain's representation of space \citep{hubel1959receptive, o1971hippocampus}. This focus on non-temporal research questions was probably supported, in particular before the predictive revolution, by an experimental view that a stimulus is an impulse which causes measurable short-lived neuronal perturbations. In this view, such perturbations are analysed and interpreted as a system response without considering the possibility that the response may be part of some underlying, ongoing dynamics. More recently, it became more evident that spontaneous brain activity may be crucial for understanding brain function \citep{fox2005human, deco2011emerging, kenet2003spontaneously}.

What does it mean to refer to neuronal activity as sequential? In the most common sense of the word, a sequence is usually understood as the serial succession of discrete elements or states. Likewise, when thinking of sequences, most people intuitively think of examples like 'A, B, C, ...' or '1, 2, 3, ...'. However, when extending this discrete concept to neuronal sequences, there are only few compelling examples where spike activity is readily interpretable as a discrete sequence, like the 'domino-chain' activation observed in the birdbrain nucleus HVC \citep{hahnloser2002ultra}. However, it is reasonable to assume that sequential representations are used in many brain areas, especially those directly connected to the hippocampus. Hippocampal-cortical interactions are characterised by so-called sharp wave ripples \citep{buzsaki2015hippocampal}, which are effectively compressed spike sequences. Recent findings suggest that other cortical areas connected to the hippocampus use grid-cell like representations similar to space representation in the hippocampus \citep{constantinescu2016organizing, stachenfeld2017hippocampus}. This suggests that at least areas connected to the hippocampus are able to decode spike sequences.

\begin{figure}[t]
\begin{center}
\includegraphics[width=13cm]{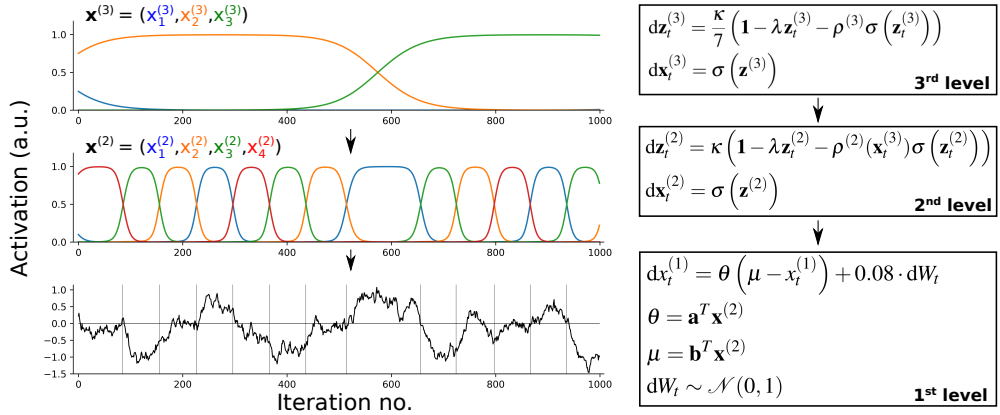}
\end{center}
\caption{Illustration of hierarchical continuodiscrete dynamics. Slowly changing dynamics at the top level parametrize the sequence of states of the faster changing second-level dynamics $z^{(2)}$. As the dynamics of variables $\vec{x}^{(i)}$ iterates between the binary states ``on'' and ``off'', their behaviour constitutes continuodiscrete WLC dynamics. At around iteration step $600$, the green unit at the third level (element $\vec{x}^{(3)}_3$) becomes active, which changes the second-level sequential dynamics from red$\rightarrow$green$\rightarrow$orange$\rightarrow$blue$\rightarrow$red to red$\rightarrow$blue$\rightarrow$green$\rightarrow$orange$\rightarrow$red. This is achieved by parametrization of the second-level connectivity matrix $\rho^{(2)}$ based on the activity on the third level. The second-level dynamics in turn parametrize an Ornstein-Uhlenbeck process, where the activity of the second-level units drives the process towards the attractors specified in vector $\vect{b}=(0.6,0,-1,-0.3)^T$. The rate of change per unit time is parametrized by vector $\vect{a}=(1,0.5,1.2,0.8)^T$. $\sigma$ is the softmax function which is applied elementwise. \textbf{1} denotes a vector of ones. $\kappa =2$, $\lambda=1/8$. Grey vertical lines in the first level mark the moment of change between states in the second level. This hierarchical parametrization of sequences is similar to the approach in \citep{kiebel2009recognizing}.}\label{continuodiscrete_trajectory}
\end{figure}

To close the gap between, on one side, the seemingly continuous dynamics in our environment, our sensory input and our own movements and, on the other side, discrete sequences, one can use so-called winnerless competition (WLC) dynamics \citep{rabinovich2000dynamical, afraimovich2004heteroclinic, rabinovich2008transient}. WLC dynamics have been used to characterise sequential neuronal activity as a succession of discrete-like states which serve as waypoints for the continuous evolution of neuronal dynamics and cognitive states. In this paper we introduce the similar concept of \textit{continuodiscrete trajectories}, which are characterised by concrete states (fixed points) which are connected by continuous trajectories. Continuodiscrete trajectories can exhibit WLC dynamics, while they can also remain in one state indefinitely. The concept of continuodiscrete dynamics is schematically illustrated in Figure \ref{continuodiscrete_trajectory}. Such an approach makes intuitive sense for describing behaviour, as many movement sequences are structured in this fashion. One illustrative example is dancing, which can be described as a succession of postures and steps, which serve the function of discrete waypoints for the choreography. These discrete sequence elements are then ``connected'' by dynamic body movements. Another intuitive example is speech, which can, at a high level, be modelled as a discrete sequence of words. At lower levels, speech can be expressed as continuous dynamics (vocal tract movements) between specific postures of the vocal tract (discrete waypoints), see e.g. \citep{birkholz2010model}.
For the brain, representing continuodiscrete trajectories seems to combine the best of two worlds: Firstly, the representation of discrete-like waypoints forms the basis for the generalisation and categorisation of the sequence. For example, for the categorisation of a specific movement sequence, it is not necessary to consider all the details of the sensory input, as it is sufficient to categorise the sequence type (dancing, walking, running) by recognising the sequence of discrete-like waypoints, as e.g. in \citep{giese2003cognitive}.  Secondly, the brain requires a way of representing continuous dynamics to not miss important details. This is because key information can only be inferred by subtle variations within a sequence, as is often the case in our environment. For instance, when someone is talking to us, most of the speech content is represented by discrete-like waypoints that describe vocal tract movements, and body movements like gestures. Additionally however, there are subtle variations in the exact expression of the discrete-like waypoints which let us infer about otherwise hidden states like the emotional state of the speaker \citep{birkholz2010model, kotz2003lateralization, schmidt2006movement}. In other words, representing sensory input as continuodiscrete trajectories enables the recognition of invariances of the underlying movements without losing fine-grained details.

There is growing evidence that sequences with discrete-like states are a fundamental feature of cognitive and perceptual representations, e.g. \citep{seidemann1996simultaneously, jones2007natural}. This feature may be at the heart of several findings in the cognitive sciences which suggest that human perception is chunked into discrete states, see \citep{vanrullen2003perception} for some insightful examples. Assuming that the brain uses some form of continuodiscrete trajectories to represent sensory input, the next question is how this can be expressed mathematically.

\section{Artificial Neural Networks as Sequence Generators}
Artificial Neural Networks generally fall into one of two categories: Feed-forward networks and recurrent networks. The distinguishing difference between these two approaches is their connectivity architecture. In recurrent neural networks (RNNs), neurons (units) maintain connections back to themselves, either directly or indirectly via other neurons. Consequently, a unit’s activity in a recurrent network has an effect on its activity at some later time point, even in the absence of time-dependent parameters like refractory time. This introduces an intrinsic time-dependency, which enables recurrent networks to deal well with temporally structured data. Conversely, activity of a unit in a feed-forward network affects only subsequent units, and does not influence its own activity at some later point, unless this is specifically modelled with time-dependent parameters. Therefore, it is reasonable to assume that feed-forward networks are generally more suitable for problems that do not exhibit an explicit temporal nature, like image recognition. Hence feed-forward networks may in general not be adequate for sequence generation, unless they are explicitly modelled to do so (see next section for a prominent example), while recurrent models of neuronal activity are either sequence generators by design, or can easily be used for sequence generation \citep{breakspear2017dynamic, long2010review, gros2009cognitive}. In what follows, we will briefly describe various dynamical principles implemented in artificial feed-forward and recurrent neural networks that have been proposed to model sequential neuronal activity. For RNNs, we will furthermore discriminate between network architectures which generate random sequences implicitly as a side-effect of their recurrent connectivities, and architectures which are explicitly designed to generate a sequential activation of their constituents.

\subsection{Feed-Forward Neural Networks}
A specific arrangement of neurons in a feed-forward chain can be exploited for sequence generation. Such arrangements, called \textit{synfire chains}, have been argued to be possibly the basis of some neuroscientific observations of sequential activity \citep{abeles1991corticonics, diesmann1999stable, ikegaya2004synfire}.
\subsubsection{Synfire Chains}
\begin{wrapfigure}[]{R}{0.65\textwidth}\begin{center}\includegraphics[width=0.37\textwidth]{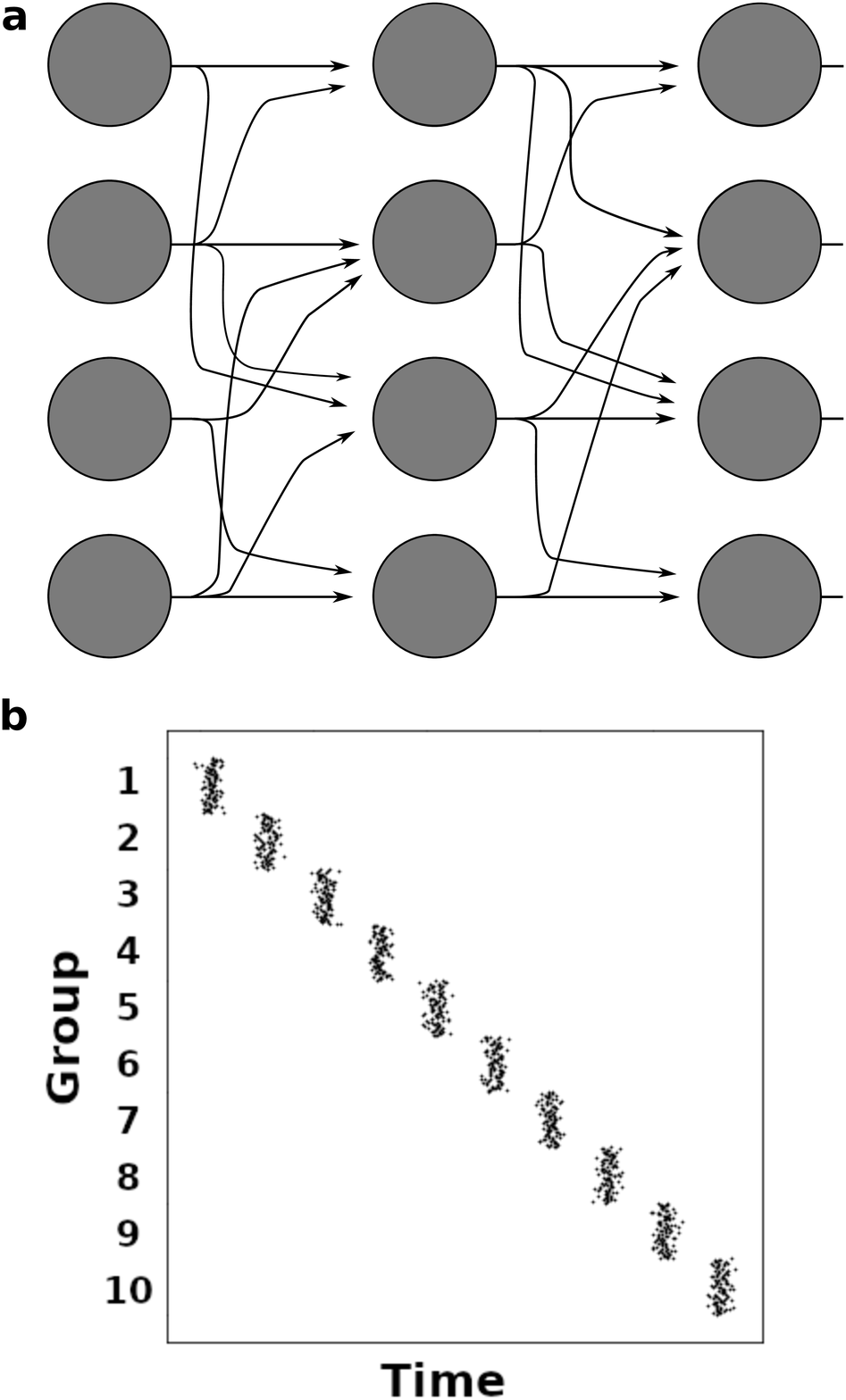}\end{center}
\caption{\textbf{a} Illustration of a synfire chain. Arrows indicate excitatory connections. \textbf{b} Spiking histogram of neurons in a synfire chain with ten nodes of 100 neurons each. The average time interval between the firing of two adjacent groups corresponds to one synaptic delay. }\label{synfire_chains_text}
\end{wrapfigure}

Synfire chains are concatenated groups of excitatory neurons with convergent-divergent feed-forward connectivity, as illustrated in Fig. \ref{synfire_chains_text} (a) \citep{abeles1991corticonics, diesmann1999stable}. Synchronous activation of one group leads to the activation of the subsequent group in the chain after one synaptic delay. It has been shown that the only stable operating mode in synfire chains is the synchronous mode where all neurons of a group spike in synchrony \citep{litvak2003transmission}. Synfire chains create a temporally highly precise sequence of firing neuronal clusters \citep{abeles1991corticonics, diesmann1999stable}. Such temporally precise sequences have been observed in slices of the mouse primary visual cortex and in V1 of anaesthetised cats \citep{ikegaya2004synfire}, as well as in the HVC song nucleus of the bird brain during song production \citep{hahnloser2002ultra, long2010support}. Synfire chains might be one explanation for these observations. A striking mismatch to neuronal networks in the brain however is the absence of recurrent connections in the synfire chain's feed-forward architecture. Modelling studies have shown that sequential activation similar to synfire chain activity can be achieved by changing a small fraction of the connectivities in a random neural network \citep{rajan2016recurrent, chenkov2017memory}. This was used to implement working memory \citep{rajan2016recurrent} or give a possible explanation for the occurrence of memory replay after one-shot learning \citep{chenkov2017memory}. Such internally generated sequences have been proposed as a mechanism for memory consolidation, among other things (see \citep{pezzulo2014internally} for a review).

\subsection{Recurrent Neural Networks}
Recurrent Neural Networks (RNN) are an important part of neuroscience, as most brain regions, and especially the cortex, are recurrently connected internally. The current role of RNNs in research is diverse, and theoretical neuroscience employs them in various different approaches. RNNs have been used to replicate findings from behavioural and imaging studies \citep{enel2016reservoir, rajan2016recurrent, laje2013robust, kim2019simple}, and some interesting research is dedicated to the evolution of neurobiologically sound network learning rules \citep{miconi2017biologically, memmesheimer2014learning}. Often, such learning rules are either biologically more plausible versions of algorithms already known from machine learning, or their actual occurrence in biological systems remains yet to be shown. Another stream of research investigates the capabilities and characteristics of networks which learn based on biologically well-established concepts like Hebbian plasticity and spike-time dependent plasticity \citep{sprekeler2007slowness, legenstein2010reward, chenkov2017memory}. A review of the current state of learning in biologically plausible spiking neural networks can be found in \citep{taherkhani2020review}.

\subsubsection{Limit Cycles}
\begin{figure}[h!]
\begin{center}
\includegraphics[width=12cm]{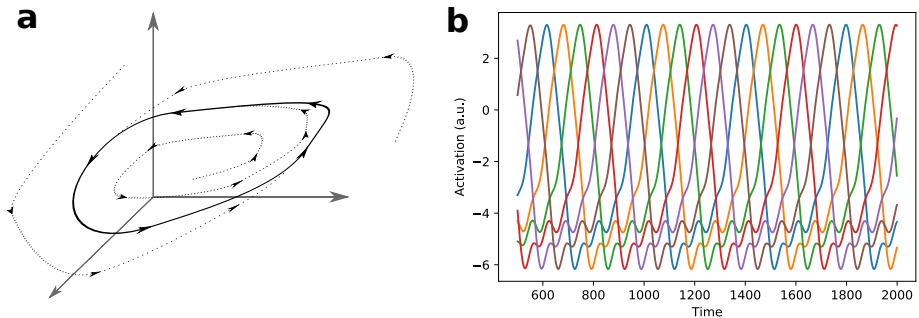}
\end{center}
\caption{Two different representations of limit cycles. \textbf{a} A Limit cycle in three-dimensional phase space. The dimensions of the phase space can be interpreted as the firing rate of single units. \textbf{b} Representation of a six-dimensional limit cycle as alternating activations of six different units.}\label{LC}
\end{figure}

Limit cycles are important objects in dynamic systems theory, and they occur practically in every physical domain \citep{strogatz2018nonlinear}. While limit cycles are continuous trajectories in phase space and do therefore not fall within the realm of continuodiscrete dynamics, experimentally they can be barely distinguishable from continuodiscrete trajectories of high temporal precision. We will therefore briefly discuss limit cycles and their application in current neuroscience research.

A limit cycle is a closed trajectory, with fixed period and amplitude, in the phase-space of a dynamical system (Fig. \ref{LC}). Nearby trajectories are either attracted towards or repelled from a limit cycle. Limit cycles occur frequently in biological and other dynamical systems, e.g. the beating of the heart, or the periodic firing of a pacemaker neuron are regulated by limit cycles \citep{strogatz2018nonlinear}. Here, we are interested in limit cycles in the phase space of a neuronal network. Limit cycles can occur in nonlinear, i.e. recurrent, neural networks \citep{berry2006structure, jouffroy2007design}. Limit cycles by definition have smooth, continuous phase-space trajectories. They are of great interest to theoretical neuroscience, as periodic spiking activity can be modelled by limit cycles, both on single-cell level \citep{izhikevich2007dynamical} and population level \citep{mi2017synaptic}. They also play an important role in the emulation of human motion in robotics, where a study conducted on human participants reported that there are two different types of human motion \citep{schaal2007dynamics}. The first type is rhythmic motion, like the pedalling on a bike, or the rhythmic step-by-step alternation of the feet during walking. The second type is called ``motion strokes'' which comprises directed non-rhythmic motion, like reaching for a glass of water. These two types of motion can be modelled by the unifying approach of \textit{dynamic motion primitives} (DMPs) \citep{schaal2007dynamics}. The main idea of DMPs is that the limbs move as if they were pulled towards an attractor state. In the case of rhythmic motion, the attractor is given by a limit cycle, while in the case of motion strokes the attractor is a discrete point in space. This suggests that sequences and limit cycles serve similar purposes.

\subsubsection{Heteroclinic Trajectories}
\begin{figure}[h!]
\begin{center}
\includegraphics[width=0.8\textwidth]{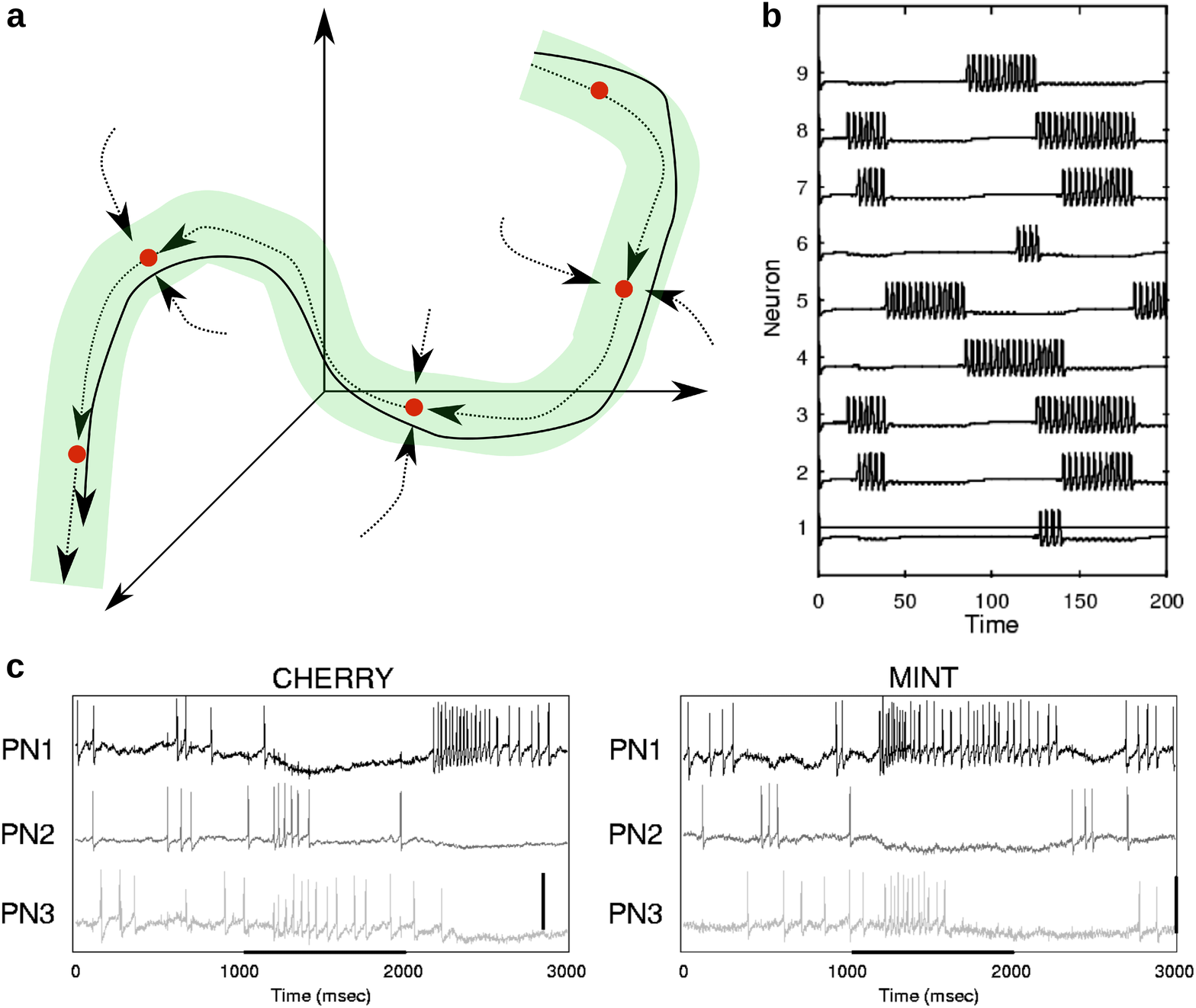}
\end{center}
\caption{\textbf{a} Illustration of a heteroclinic trajectory (solid line) in three-dimensional phase space. Dashed lines indicate invariant manifolds between saddle states, i.e. het\-eroclin\-ic connections. The green tube illustrates a Stable Het\-eroclin\-ic Channel. All het\-eroclin\-ic trajectories originating in the SHC will remain inside of it. \textbf{b} Simulation of an SHC-trajectory based on Lotka-Volterra dynamics. \textbf{c} Neuronal responses to odour representation in the locust brain. Figures \textbf{b} and \textbf{c} are adapted with permission from \citep{rabinovich2001dynamical} (\href{https://dx.doi.org/10.1103/PhysRevLett.87.068102}{DOI}). Copyright (2001) by the American Physical Society.}\label{SHC}
\end{figure}
Another approach to model sequential neuronal dynamics are het\-eroclin\-ic networks \citep{ashwin2005nonlinear, rabinovich2008transient}. A het\-eroclin\-ic network is a dynamic system with semi-stable states (saddle points) which are connected by phase-space trajectories. 
Symmetric networks of coupled oscillators exhibit useful mathematical prop\-er\-ties, including the occurrence of robust het\-eroclin\-ic cycles \citep{ashwin1992dynamics}. This is interesting for theories of neuronal networks as the activity of single neurons is well described by oscillator models, and synchro\-nized activity of neurons can be described by coupled oscillator models \citep{ashwin2007dynamics}. Coupled oscillator networks were observed to exhibit het\-eroclin\-ic behaviour. This behaviour is characterised by phase space trajectories along the heteroclinic connections of the network (Fig. \ref{SHC}). This itineration can be guided by external input, where the travelled trajectory discriminates between different inputs \citep{ashwin2007dynamics}. It was furthermore shown that such het\-eroclin\-ic networks generally possess the ability to perform computational tasks \citep{ashwin2005discrete}, and that it is possible to implement any logic operation within such a network \citep{neves2012computation}. Closely related to the dynamic alteration between states in a het\-eroclin\-ic network is the previously introduced concept of winnerless competition (WLC) \citep{rabinovich2000dynamical, afraimovich2004heteroclinic, rabinovich2008transient} (see Sec. \ref{sec:what_are_sequences}). This concept can be viewed as a counterpoint to the well-known winner-take-all dynamics of sensory coding, as is for example observed in Hopfield networks \citep{hopfield1982neural}. WLC was proposed as a general information processing principle for dynamical networks and is characterised by dynamic switching between network states, where the switching behaviour is based on external input \citep{afraimovich2004heteroclinic}. Importantly, the travelled trajectory identifies the received input, while any single state of the trajectory generally does not, see for example \citep{neves2012computation}. In phase space representation, WLC can be achieved by open or closed sequences of het\-eroclin\-ically concatenated saddle points. Such sequences are termed stable het\-eroclin\-ic sequences (SHS) if the het\-eroclin\-ic connections are dissipative, i.e. when a trajectory starting in a neighbourhood close to the sequence remains close \citep{afraimovich2004origin}. The application of SHS to real physical dynamical systems, like networks of biological neurons, is however limited as many kinds of external and internal perturbations can destroy an SHS. If trajectories of an SHS are preserved, the sequence and its surrounding stable neighbourhood are called a Stable Het\-eroclin\-ic Channel (SHC) (see Fig. \ref{SHC}) \citep{rabinovich2006generation}. A simple implementation of SHCs is based on the generalised Lotka-Volterra equations \citep{bick2010occurrence}, which are a type of recurrent neural network implicitly implementing the WLC concept. The temporal precision of a system that evolves along an SHC is defined by the noise level as well as the eigenvalues of the invariant directions of the saddle points. Therefore, sequences along heteroclinic trajectories are reproducible although the exact timing of the sequence elements may be subject to fluctuation.

In a similar approach, recent theoretical work on the behaviour of RNNs has introduced the concept of excitable network attractors, which are characterised by stable states of a system connected by excitable connections \citep{ceni2019interpreting}. The conceptual idea of orbits between fixed points may further be implemented in different ways. For instance, transient activation of neuronal clusters can be achieved by introducing bifurcations such that the currently stable fixed point becomes unstable, thus forcing the system to approach a new stable state \citep{gros2007neural}.
\section{Recognition of sequences}
\label{sec:seq_generators_gen_models}
Although neuronal sequence models such as the ones introduced in the preceding sections have been used to model experimentally observed neuronal data, these models by themselves do not explain how predictions are formed about the future trajectory of a sequence. Rather, neuronal sequence models like synfire chains or SHCs only describe how trajectories travel from one waypoint to another, in an autonomous fashion. To take the example of song production and recognition in songbirds, if we have a sequential model of birdsong generation, e.g. the song of a Northern cardinal, this is not be sufficient to model or explain how the listening female bird recognises the ongoing song \citep{yildiz2011hierarchical}. To model recognition, one needs a predictive model which compares the listening female's ongoing expectations of the male's song with the auditory input. One can generalise this point to any sensory input: To recognise and predict the observed sensory input caused by movement of others (or oneself), one needs a recognition model with expectations about the expected, underlying movement sequence that caused the sensory input. 
One way to do this is to combine sequential models with the Bayesian brain hypothesis. The idea is to use sequential models as generative models of sensory input so that one can derive Bayesian recognition models for sequentially generated sensory input. A simple example of such a scheme is provided in \citep{bitzer2012recognizing}, where RNNs are used as a generative model such that model inversion provides for an online recognition model. As shown in \citep{friston2011action}, one can also place such a generative model into the active inference framework to derive a model that not only recognises sequential movements from visual input but also generates sequential movement patterns. Such models are not only interesting from a cognitive neuroscience perspective but also point at a shared interest with the field of artificial intelligence and specifically machine learning, of better understanding the nature of generative models employed by the brain. In other words, both cognitive neuroscience and artificial intelligence strive to find a mechanistic understanding of how sensory input caused by others can be recognised by an artificial or a biological agent. In the following, we will review how both fields seem to converge on the conceptual idea that such models should be sequential and hierarchical. 

\subsection{Sequence recognition in machine learning}
\label{sec:sequences_in_ML}
Given that much of our sensory input can be modelled as continuodiscrete sequences, the question is how an artificial agent can recognise such sequences under the Bayesian brain hypothesis. Sequential inference methods have a long history in both machine learning and neuroscience. The most widely used models for discrete sequence generation are hidden Markov models (HMM) and their time-dependent gen\-er\-al\-i\-sa\-tion, hidden semi-Markov models (HSMM) \citep{yu2015hidden}. In particular, HMMs and HSMM are a standard tool in a wide range of applications concerned with recognising human movements, e.g. in speech recognition \citep{liu2018structured,zen2004hidden,deng2006structured} and activity recognition \citep{duong2005activity}.

Similar to HSMMs, artificial RNNs are actively used in machine learning for classifying and predicting time series data. When training a generic RNN for prediction and classification of time series data, one faces various challenges, most notably incorporating information about long-term dependencies in the data. To address these dependencies, specific RNN architectures have been proposed, such as \textit{long-short term memory} (LSTM) networks \citep{gers1999learning} and \textit{gate recurrent units} (GRU) \citep{chung2014empirical}. In a common LSTM network, additionally to the output variable, the network computes an internal variable called the memory cell, which it passes on to itself upon each iteration. This endows the network with high flexibility. LSTM networks belong to the most successful and most widely applied RNN architectures, with applications in virtually every field involving time-series data, or any data structure with long-range dependencies \citep{yu2019review, lecun2015deep}. Another RNN approach is \textit{reservoir computing} (RC), which started with the development of echo-state networks and spiking liquid state machines in the early 2000s \citep{lukovsevivcius2012reservoir, jaeger2001echo, maass2002real}. In RC, sequential input is fed to one or more input neurons. Those neurons are connected with a \textit{reservoir} of randomly connected neurons, which in turn are connected to one or more output neurons. During training of a reservoir computing network only the output weights are adjusted. RC networks have successfully been applied in almost every field of machine learning and data science, such as speech recognition, handwriting recognition, robot motor control and financial forecasting \citep{lukovsevivcius2012reservoir, tanaka2019recent}. 

In summary, machine learning approaches for sequence prediction and classification are developing in the direction of models capable of representing sequences. The existence of different long-term dependencies does furthermore suggest that sequences can be present over a mixture of time scales, see below.

Although machine recognition models with intrinsic sequential dynamics are commonplace in computer science and AI research, most existing models used in machine learning are not neurobiologically plausible mainly for two reasons: either they are purely discrete (HSMM), or use complex non-local learning schemes (e.g. backpropagation signals over multiple layers), although there is research on neurobiologically sound learning paradigms for RNNs \citep{miconi2017biologically, taherkhani2020review}. One alternative is to use neurobiologically plausible sequence generation models, which can act as generative models of the causal dynamic relationships in the environment. A natural application would be the development of models of the recognition process in terms of Bayesian inference \citep{bitzer2012recognizing}, and more specifically in terms of variational inference \citep{friston2006free, daunizeau2009variational}.

\subsection{Biological and Artificial Inferential Hierarchies}

In neuroscience and the cognitive sciences, the brain is often viewed as a hierarchical system, where a functional hierarchy can be mapped to the structural hierarchy of the cortex \citep{badre2008cognitive, koechlin2003architecture, kiebel2008hierarchy}. The best example of such a hierarchical organisation is the visual system, for which the existence of both a functional and an equivalent structural hierarchy is most unambiguously established \citep{felleman1991distributed}. Cells in lower levels of the hierarchy encode simple features and have smaller receptive fields than cells further up the hierarchy, which in turn posses larger receptive fields and encode more complex patterns by integrating information from lower levels \citep{hubel1959receptive, zeki1988functional, giese2003cognitive}. This functional hierarchy is mediated by an asymmetry of recurrent connectivity in the visual stream \citep{zeki1988functional, sherman1998actions}.
Hierarchical organisation of a biological neuronal system appears furthermore to occur naturally as a result of connection costs between neural ensembles, which has been shown in a modelling study with artificial neurons \citep{mengistu2016evolutionary}. Hierarchies also play an important role in cognitive neuroscience as most if not all types of behaviour as well as cognitive processes can be described in a hierarchical fashion. For example, the action of making a cup of tea can be considered a high-order goal in a hierarchy with various subgoals which are temporally less extended and less abstract. In the given example, these subgoals can be: (i) putting a teabag into a pot, (ii) pouring hot water into the pot and (iii) pouring tea into a cup (example adopted from \citep{botvinick2007multilevel}). 

Importantly, all theories of cortical hierarchies of function share the common assumption that primary sensory regions are at the bottom of the hierarchy and are influenced by higher order association cortices. This principle has been conceptualized as a 'hierarchy of time scales' \citep{kiebel2008hierarchy, hasson2008hierarchy, koechlin2003architecture, badre2008cognitive}. In a hierarchy of time scales the lower levels of the hierarchy represent relatively quickly changing dynamics, which encode the fast features of sensory input. These levels can therefore be identified with early sensory regions. Levels further up the hierarchy code for more general characteristics of the environment and inner cognitive processes, which generally change slowly \citep{hasson2008hierarchy, koechlin2003architecture, badre2008cognitive}.

The assumption that higher order cortical processes encode more general, slowly varying features is interesting when considering recent advances in the field of neuronal computation, where it was shown that spike-time dependent plasticity enables neurons to extract the slow features of a whitened input signal \citep{sprekeler2007slowness}. This is a crucial step in the machine learning algorithm of ``slow feature analysis’’, which extracts the slow dynamics from an input signal \citep{wiskott2002slow}. The hypothesis of a hierarchy of time scales is furthermore appealing as it relies on the rather general concept of time as compared to concepts relating to the abstractness of representations which may be well defined in behavioural experiments but are less clear in everyday situations. We assume that the brain aims to mimic the sensory input it receives in order to predict the dynamic evolution of its environment. Dynamical processes in the environment do however evolve on different time scales, where quickly changing dynamics are often ``enslaved’’ to the slower dynamics. For example, the observation of a running person is governed by the rather slowly changing dynamics of her general direction, while these slow dynamics are superposed by the quickly changing dynamics of her limbs, and the evoked excitation of retinal cells are in turn superposed by yet faster visual fluctuations of visual input. In order to make sense of the world and its surroundings, an agent therefore needs to account for dynamics at different time scales, and be able to relate their respective influences on each other.

Additional experimental support for the hypothesis of a  hierarchy of time scales has been reported in several neuroimaging studies. In a meta-analysis of six studies recording single-unit activity in the primate brain during behavioural tasks, the time scale of brain regions was inferred on the basis of the autocorrelation function of spike counts for each area \citep{murray2014hierarchy}. The results provide evidence for an organisation into a hierarchy of time scales, which was consistent with the anatomical hierarchy inferred from long-range cortical projections. The sensory cortex consistently showed short time scales, while the parietal association cortex exhibited intermediate, and medial prefrontal areas and anterior cingulate cortex (ACC) long time scales \citep{murray2014hierarchy}. Furthermore, in an fMRI study, the temporal receptive windows (TRW) of several brain regions in the human brain were investigated \citep{hasson2008hierarchy}. The TRW of an area is the time-interval over which the region ``integrates’’ incoming information, in order to extract meaning over a specific temporal scale. It was found that regions such as the primary visual cortex exhibited rather short TRW, while high order regions exhibited intermediate to long TRW \citep{hasson2008hierarchy}.

The importance of cognition at different time scales is also illustrated by various computational modelling studies. In one study, robots were endowed with a layered neural network whose parameters were let free to evolve over time to optimise performance during a navigation task \citep{nolfi2002evolving}. After some time, the robots had evolved neural assemblies with clearly distinct time scales: one assembly had assumed a quickly changing, short time scale associated with immediate sensory input while  another assembly had adopted a long time scale, as the navigation task required integration of information over an extended period of time. Another modelling study showed that robots with neuronal populations of strongly differing time-constants performed their tasks significantly better than when endowed only with units of approximately identical time-constants \citep{yamashita2008emergence}. In \citep{botvinick2007multilevel} it was shown that, after learning, a neural network with a structural hierarchy similar to one proposed for the frontal cortex had organised in such a way that high level units coded for temporal context while low-level units encoded fast responses similar to the role assigned to sensory and motor regions in theories of hierarchical cortical processing.

Similarly, a hierarchically structured RNN combined with a model of songbirds' vocal tract dynamics was used in \citep{yildiz2011hierarchical} to model the process of birdsong generation and inference in songbirds. The system consisted of three levels, each of which governed by the sequential dynamics of an RNN following a limit cycle. The sequential dynamics were then influenced both by top-down predictions from the higher level, and prediction errors from the level below. The model was then used for birdsong recognition. In another study, the same concept was used for recognition of human speech \citep{yildiz2013birdsong}. The resulting inference scheme was able to recognise spoken words, even under adversarial circumstances like accelerated speech, since it entails parameters which are adaptable in an online fashion during the recognition process. The same principle can also be applied to very different types of input, see \citep{rivera2015modelling} for an example in insect olfaction.

Not surprisingly, the importance of hierarchies of time scales is well established within the machine learning community \citep{malhotra2015long,el1996hierarchical}. Current state-of-the-art RNN architectures used for prediction and classification of complex time series data are based on recurrent network units organised in temporal hierarchies. Notable examples are the clockwork RNN \citep{koutnik2014clockwork}, gated feedback RNN \cite{chung2015gated}, hierarchical multi-scale RNN \cite{chung2016hierarchical}, fast-slow RNN \cite{mujika2017fast}, and a higher order RNN architecture \cite{soltani2016higher}. These modern RNN architectures have found various applications in motion classification \cite{yan2018hierarchical,neverova2016learning}, speech synthesis \citep{zhang2018high,achanta2017deep,wu2016investigating}, recognition \citep{chan2016listen}, and other related areas \cite{kurata2017language,krause2017hierarchical,liu2015multi}. Importantly, these applications of hierarchical RNN architectures further confirm the relevance of hierarchically organised sequence generators for capturing complex dynamics in our everyday environments.

\section{Conclusion}
\label{sec:Discussion}
In summary, there is a remarkable body of evidence from both experimental and computational neuroscience that the brain's causal model of the environment is structured as a hierarchy of time scales suitable for generating and recognising complex sequential activation patterns. In addition, a steadily increasing amount of empirical evidence suggests that the brain integrates experience and current sensory information to update its causal model of the world. Here we reviewed various models of sequential neuronal activity and work demonstrating how robust inference can be performed using such models. We argue that generative models of sequences can model and predict sequential processes in real-world sensory data, especially if that sensory data were generated by human movements. Inference based on these sequential, generative models in principle leads to robust belief formation and the adaptation to changing dynamical conditions in the environment.

Currently, both computational neuroscience and machine learning are developing in parallel, with many intersections, in an effort to understand key principles of prediction of, and action upon, complex time series. Clearly, the effort of machine learning research to represent and predict time series data in various technological domains, and the effort of neuroscientists to describe the functional role of sequential neuronal activity in the brain, are tightly linked. Both fields are constantly evolving and influencing each other with important empirical data and theoretical concepts. In particular, it is reasonable to assume that the recognition and prediction of human intentions and how humans employ cognitive control, in an online fashion, based on observed human movements, will be a converging point for future developments both in machine learning and cognitive neuroscience.  

\section{Author Contributions}
DM and SK both contributed to the conception of the study. SF wrote the manuscript, with contributions by DM and SK. 

\section{Conflict of Interest Statement}
The authors declare no conflicts of professional, personal, or financial interest.

\section{Funding Disclosure}
Funded by the German Research Foundation (DFG, Deutsche Forschungsgemeinschaft), SFB 940/2, project A9, and as part of Germany’s Excellence Strategy – EXC 2050/1 – Project ID 390696704 – Cluster of Excellence “Centre for Tactile Internet with Human-in-the-Loop” (CeTI) of Technische Universität Dresden.

\bibliographystyle{apalike}
\bibliography{references}

\end{document}